\documentclass{osa-article}

\journal{boe}

\articletype{Research Article}

\begin{document}



\title{Deep-3D Microscope: 3D volumetric microscopy of thick scattering samples using a wide-field microscope and machine learning}

\author{Bowen Li,\authormark{1} Shiyu Tan,\authormark{2} Jiuyang Dong,\authormark{3} Xiaocong Lian,\authormark{1} Yongbing Zhang,\authormark{4}
Xiangyang Ji,\authormark{1,*} Ashok Veeraraghavan\authormark{2,*}}

\address{
\authormark{1}Department of Automation, Tsinghua University, Beijing, China\\
\authormark{2}Department of Electrical and Computer Engineering, Rice University, Houston, TX 77005, USA\\
\authormark{3}Tsinghua Shenzhen International Graduate School, Shenzhen, China\\
\authormark{4}Harbin Institute of Technology (Shenzhen), Shenzhen, China\\

\email{\authormark{*}vashok@rice.edu, xyji@tsinghua.edu.cn}} 


\begin{abstract}
Confocal microscopy is the standard approach for obtaining volumetric images of a sample with high axial and lateral resolution, especially when dealing with scattering samples.
Unfortunately, a confocal microscope is quite expensive compared to traditional microscopes.
In addition, the point scanning in a confocal leads to slow imaging speed and photobleaching due to the high dose of laser energy.
In this paper, we demonstrate how the advances in machine learning can be exploited to "teach" a traditional wide-field microscope, one that's available in every lab, into producing 3D volumetric images like a confocal.
The key idea is to obtain multiple images with different focus settings using a wide-field microscope and use a 3D Generative Adversarial Network (GAN) based neural network to learn the mapping between the blurry low-contrast image stack obtained using wide-field and the sharp, high-contrast images obtained using a confocal.
After training the network with widefield-confocal image pairs, the network can reliably and accurately reconstruct 3D volumetric images that rival confocal in terms of its lateral resolution, z-sectioning and image contrast.
Our experimental results demonstrate generalization ability to handle unseen data, stability in the reconstruction results, high spatial resolution even when imaging thick ($\sim40$ microns) highly-scattering samples.
We believe that such learning-based-microscopes have the potential to bring confocal quality imaging to every lab that has a wide-field microscope.
\end{abstract}


\section{Introduction}
\label{sec:intro}
High-throughput, high-resolution, high-contrast microscopy techniques, that do not damage tissue are critical for multiple domains including scientific imaging, pathology, medical imaging, and in-vivo imaging.
The current workhorse of microscopy is a wide-field microscope and every science lab, pathologist's office and hospital/clinic in every corner of the globe likely has access to one.
While wide-field microscopes have truly been democratized, they are for the most part only suited to image the surface of thin samples. 3D volumetric imaging, especially with scattering tissue samples is a rapidly growing need that wide-field microscopes cannot address.

Existing techniques for 3D volumetric imaging in scattering samples such as confocal microscopy \cite{wilson1990confocal,pawley2006handbook}, two-photon microscopy\cite{helmchen2005deep, so2000two,park2015high,lu2017video,papadopoulos2020dynamic,zong2021miniature}, and light-sheet microscopy\cite{chen2014lattice, keller2008reconstruction,power2017guide,gao2012noninvasive,mcdole2018toto,yue2020long} all rely on more complex optics and illumination designs that end up being prohibitively expensive for many parts of the world. 
The question we ask in this paper is "Can the revolutionary advances in machine learning over the last decade be exploited to turn the data acquired from conventional wide-field microscopes to rival 3D volumetric data acquired using confocal microscopes --- even when imaging thick scattering samples?"\\

\noindent \textbf{AI-enhanced fluorescence microscopy.}
Machine learning and artificial intelligence have made revolutionary advances in the last decade and have completely transformed a variety of applications all the way from autonomous vehicles to medical diagnostics.
These revolutionary advances have been the result of (a) new deep neural network architectures that are highly over-parameterized, (b) datasets acquired to "teach" these networks, and (c) efficient algorithms both for training and for testing. This AI revolution has also begun to have significant impacts on microscopy, biological and scientific imaging. 
The key advantage is that, with no or little modification to conventional microscopes, AI techniques can provide additional capabilities such as high resolution\cite{rivenson2017deep,wang2019deep,zhang2020exceeding}, 
high SNR \cite{goy2018low,weigert2018content}, 
fast acquisition \cite{ouyang2018deep,nehme2020deepstorm3d,li2020fast}, 
autofocusing \cite{wu2018extended,pinkard2019deep}, 
and cross-modality \cite{ounkomol2018label,christiansen2018silico,rivenson2019virtual,wu2019three}.

Over the last few years researchers have developed early examples of deep convolutional neural networks to enhance axial resolution and imaging contrast of wide-field images\cite{Xiaoyu2018Deep,9136890,ning2020deep,lim2020three,huang2021recurrent}. Specifically,  Zhang et al.\cite{Xiaoyu2018Deep} first successfully transformed wide-field images to background reduced Structured Illumination Microscopy (SIM) images using a 2D neural network. However, their only demonstration was limited to samples with relatively simple structures. 
Although they subsequently demonstrated more complicated mice whole-brain images reconstruction in \cite{ning2020deep}, the 2D convolutional structure confined the application of the network to physically sectioned tissue samples. 
Wu et al. proposed two more sophisticated Generative Adversarial Network (GAN) based deep networks to predict confocal images from a single in-focus wide-field input\cite{wu2019three} and sparse wide-field scans\cite{huang2021recurrent}, respectively. 
However, they only testified this idea using 2.4 microns thickness BPAEC sample or a limited z-span of C-Elegans sample (about 4 microns). 

\begin{figure*}
\begin{center}
\includegraphics[width=1.0\linewidth]{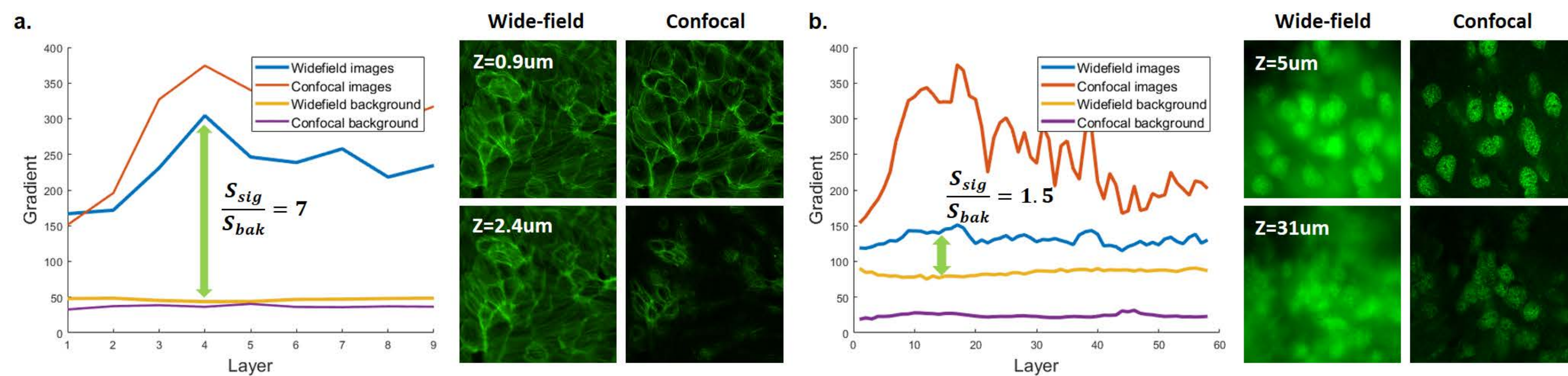}
\end{center}
\caption{
\textbf{Sharpness analysis for thin/thick samples.} 
The sharpness of the signal region and background noise at different axial depths are plotted. a)The sharpness curves for the wide-field and confocal images of a 3-microns MCF10A thin sample. b)The sharpness curves for the wide-field and confocal images of a 38-microns neuron thick sample. In the thin sample, the molecular structure is clear in the focus plane and less influenced by background noise, while the structures in the thick sample is severely deteriorated by background and scattering noise.}
\label{fig:analysis}
\end{figure*}

Here, in our paper, we take a significant leap compared to these related works.
While all of the above approaches were limited in the sample thickness/simple structures, we aim to perform true three dimensional reconstructions in thick scattering samples (demonstrating 10x thicker samples than prior work) with more complex structures. In the thick scattering sample, the images captured by a wide-field microscope are more severely deteriorated by the scattering background and noise than a thin sample. As show in Figure \ref{fig:analysis}, the contrast (sharpness ratio between signal region and background region) of wide-field images captured at different axial layers in a 38-microns neuron slice can be as much as $5\times$ lower than the contrast in a 3-microns MCF10A thin sample. Molecular structures are submerged in the noisy background of thick sample, which brings us more difficulties on wide-field to confocal predictions using deep neural networks. The key technical insight that allows us to move from thin scattering samples to thick samples, is the idea that in thick samples there is a lot of information cross-talk between and across sections (z-stack) and 2D neural networks are sub-optimal for capturing the structure of these complex interactions. 
We develop a 3D convolutional neural network structure that allows us to learn statistical relationships across the entire thick sample, allowing high-resolution, high-contrast 3D reconstructions over the entire volume.

In particular, we propose a GAN-based three-dimensional (3D) convolutional neural network (WFCON-Net) that can digitally predict confocal z-stack images from measurements of a wide-field microscope. We believe this technology will allow widely accessible wide-field microscopes to capture 3D volumetric imaging datasets of thick scattering samples -- with a quality comparable to (but slightly worse than) a confocal microscope. Our work is inspired and motivated by the recent work\cite{wu2019three} but with three significant contributions. First, we propose a 3D convolutional network to leverage the inter-layer connections other than recurrent blocks, to utilize the stronger crosstalks between different layers in thick/dense tissue samples. This strengthens the learning power to successfully recover fine structures under high magnification with stronger scattering backgrounds. Second, we add a photo-realistic VGG loss to preserve image high-frequency details. Third, we propose a 3D tailored registration technique -- point spread function (PSF) based registration to accurately align cross-modality wide-field and confocal image pairs under high noise disturbance. 
Furthermore, WFCON-Net can estimate dense confocal z-stacks from fewer wide-field z-scans, and thereby has the potential for further reducing the sample acquisition time. 
Also, we show that WFCON-Net has a good generalization ability to unseen sample data. Therefore, with the proposed method we can digitally obtain high-resolution confocal z-stacks using a typical wide-field microscope, without sacrificing the imaging depth, speed, resolution, or field of view (FOV).
In summary, by using the GAN-based 3D convolutional neural network, together with the VGG loss and the 3D tailored  registration techniques, we succeeded in recovering true 3D confocal fluorescence of thick scattering samples from enormously degraded wide-field input.

\section{Methods}
\label{sec:methods}
\begin{figure*}
\begin{center}
\includegraphics[width=1.0\linewidth]{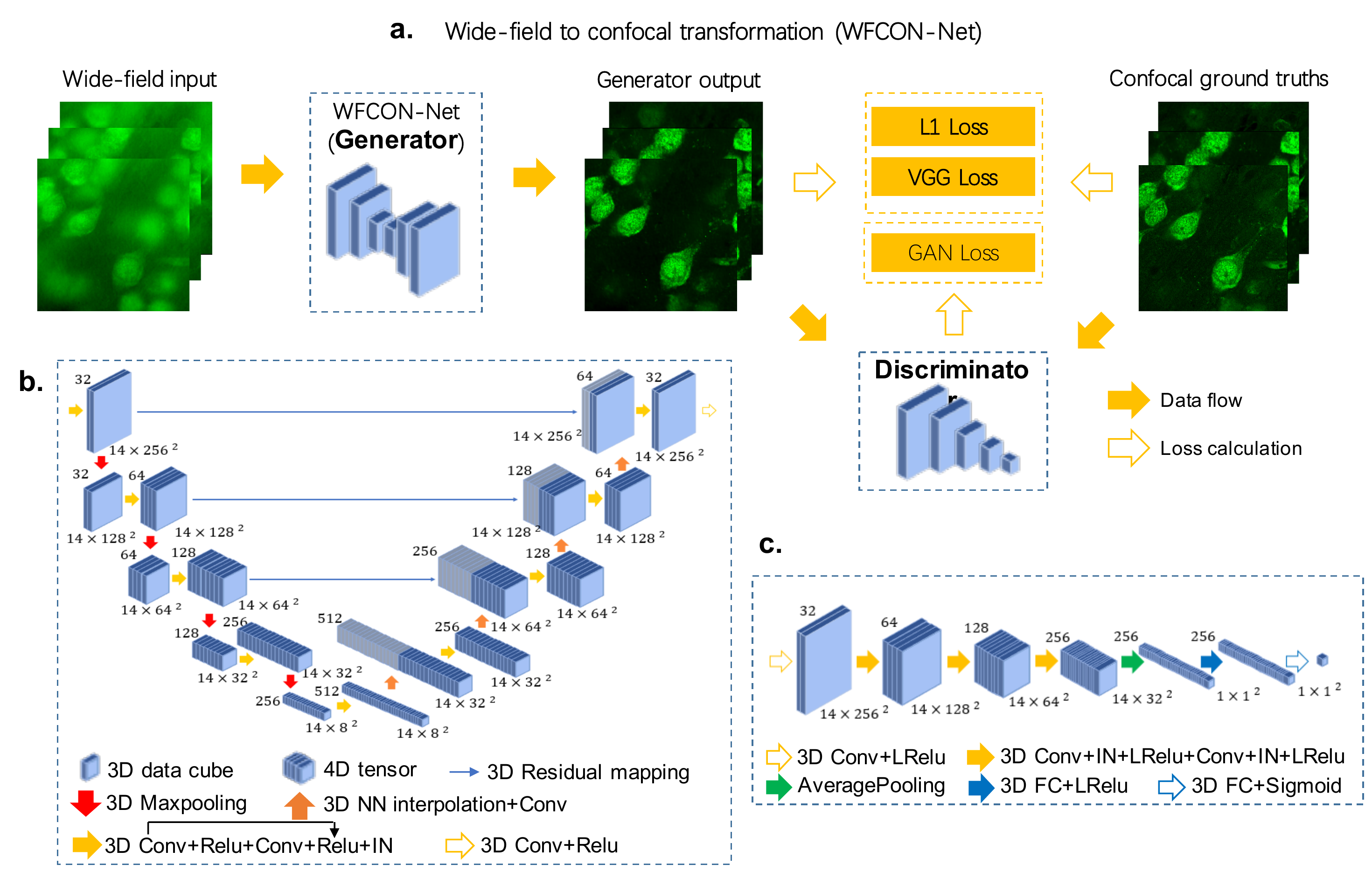}
\end{center}
\caption{
\textbf{Overview}. 
\textbf{a)}. The WFCON-Net is a 3D GAN-based deep neural network, consisting of a generator and a discriminator. 
The generator takes 3D stacked wide-field fluorescence images as input, and outputs corresponding confocal z-stack images in a single inference. 
During training, the discriminator is learned to distinguish confocal ground truths from the predictions. The use of the discriminator encourages the generator to predict confocal z-stacks with high accuracy, providing a good match to ground truth images.
\textbf{b)} and \textbf{c)}. Network architectures of generator and discriminator.}
\label{fig:network}
\end{figure*}

Consider a wide-field microscope imaging a thick scattering tissue sample. 
Typically the images obtained will suffer from low-contrast and blur associated with both in-plane and out-of-plane scattering making images (especially beyond the first 10 microns of tissue) practically un-usable.
Even so, there are significant degrees of freedom in a wide-field microscope that one can take advantage of.
The focus setting of the microscope can be slowly changed from the top to the bottom of the tissue sample to obtain a low-contrast image stack.
Regardless, each image contains a \emph{different but known} linear combination of light from the entire 3D tissue sample.
The question we ask, is whether this is sufficient information to de-multiplex and recover a sharp, high-contrast volumetric image of the sample. 
In particular, we wish to leverage deep learning techniques and use a deep generative adversarial network to learn a mapping between the blurry, low-contrast z-stack obtained using a wide-field microscope and a sharp, high contrast 3D volume imaged using a confocal microscope.\\

\noindent \textbf{WFCON-Net architecture.} 
WFCON-Net is a 3D GAN-based deep neural network, whose architecture is shown in Figure \ref{fig:network}.
The input to the network is a stack of wide-field images obtained using a wide-field microscope. 
The output of the network is the prediction of what the corresponding sharp, high-contrast 3D volumetric image obtained using a confocal microscope would be.
The network consists of two parts: a generator and a discriminator. 
The generator takes 3D stacked wide-field fluorescence images as input and outputs corresponding confocal z-stack images in a single inference.
During training, the discriminator similar to \cite{wu2019three} is learned to distinguish confocal ground truths from the predictions. The use of the discriminator encourages the generator to predict confocal z-stacks with high accuracy, together with appropriate loss function, providing a good match to ground truth images.

The generator is a modified 3D convolutional U-net \cite{cciccek20163d}, consisting of an encoder path followed by a decoder path.
The encoder path contains four down-sampling blocks: a max-pooling layer, two 3x3 3D convolutional layers, an instance normalization layer\cite{ulyanov2016instance} and a Relu activation layer\cite{glorot2011deep}.
The use of the normalization layer makes the training of thick samples with diversely distributed signals and strong scattering backgrounds stable. 
In the decoder path, the max-pooling layer is symmetrically replaced by a nearest neighbor interpolation layer followed by a convolution layer with stride $1$. 
The nearest neighbor interpolation layer, in our case, encourages the upsampling with fewer checkerboard artifacts than transpose convolutions. 
Moreover, residual mappings are performed between each convolutional layer to guarantee the gradient flow. 
All the convolution and normalization operations are implemented in a 3D manner to explore the inter-layer relation of the volumetric z-stack data. During training, the loss of the generator and the discriminator are defined as:
\begin{equation}
    \begin{split}
    &L_{loss}^G = [D(G(x))-1]^2+\lambda L_{L1}(G(x),y)+\zeta L_{VGG}(G(x),y)\\
    &L_{loss}^D = [D(G(x))]^2+[D(y)-1]^2\\ 
    \end{split}
\end{equation}
where $x$ refers to the wide-field z-stack images, $y$ refers to the corresponding sharp images captured by a confocal microscope, considered as the ground truth. $G$ and $D$ denote the generator and the discriminator. The generator loss contains a least-square GAN loss\cite{mao2017least} (with an additional L1 regularizer) and a perceptual VGG loss\cite{johnson2016perceptual}, where $\lambda$ and $\zeta$ are the corresponding weights. The use of the perceptual loss encourages high-quality, high-resolution predicted confocal images. In this paper, we set $\lambda=2$, $\zeta=0.01$ for all the experiments.\\

\noindent \textbf{Training and testing data acquisition.} 
To train/test our network, we captured 39 pairs of wide-field and confocal z-stacks images (with the lateral size of $2048\times2048$), using a developed setting of Andor Dragonfly spinning-disk confocal microscope that contains both wide-field fluorescence image capture mode and confocal fluorescence image capture mode. 
These image pairs of different regions of interest (ROIs) were randomly selected with different structures and neuron densities, covering the characteristics of different part of the tissue slice, and are split into 'training' (31 pairs) and 'testing' (8 pairs) sets. As a similar remark to \cite{cciccek20163d}, more data cannot significantly enhance reconstruction quality but at the price of computational burden. The z-stacks are scanned with a step size of $0.5\mu m$. The number of scans varies from $35$ to $76$, depending on the distribution of fluorescent signals along the z-axis. The same objective (60x/1.4NA oil, Nikon) was used for both wide-field and confocal imaging, and the resulting pixel size (in the image plane) is $108.3nm$.

During training, the z-stacks were randomly cropped into  256x256x12 3D data patches. The data patches were then augmented with random flips and rotations, and normalized to $[0,1]$ before inputting to the network. We trained our network for 6000 iterations (equivalent $\sim$ 60 epochs) using NVIDIA TitanXp GPU and it takes 3 days for training.\\

\noindent \textbf{Accurate image registration.} To ensure the reconstruction quality of thick samples under high magnification, accurate image registration is indispensable. However, severe background and decreased contrast in wide-field images make commonly used cross-correlation registration (or calculate SSIM value)  prone to error. Therefore, we proposed a new registration method tailored for 3D that calculates the PSF between confocal and wide-field image stacks, which can learn the physical connections between two stacks that is more robust to noise. Then we use this PSF to determine the lateral and axial shifts. We termed three-dimensional confocal images as $x$, wide-field images as $y$.  For simplicity, we treat the confocal images as ground-truth of sample distribution, then $y=f\ast x+n$, where $f$ is the PSF of wide-field images and noise $n$ results from non-uniform system/model errors and randomness of measurement. To robustly recover PSF, we add standard TV constraints on the gradients of the recovered PSF. Therefore, we can formulate the objective function as:
\begin{equation}
f_{opt}= \mathop{\arg\min}_{f}\lVert fx-y\rVert_2^2+\lambda \lVert \nabla f \rVert_1 +\mu \lVert 1^T f-1\rVert_2^2
\end{equation}
where the first term is the least-squares data fitting term, the second term is TV gradients penalty, the last term enforces the energy conservation constraint, i.e., $\Sigma_{m,n}f(m,n)=1$. In this experiment, $\lambda$ is set to 1 and $\mu$ is set to 10.
We use optimal first-order primal-dual framework\cite{heide2013high,chambolle2011first} to optimize this objective function to get optimal $f$:
\begin{equation}
    \begin{split}
    g_{n+1}&=Prox_{\sigma F^*}(g_n+\sigma K\bar{f_n})\\
    f_{n+1}&=Prox_{\tau G}(f_{n}+\tau K^*g_{n+1})\\
    \bar{f_n}&=f_{n+1}+\theta(f_{n+1}-f_n)
    \end{split}
\end{equation}
Where $\sigma,\theta$ and $\tau$ are hyper-parameters, $K$ is gradient operator, $*$ denotes the convex conjugate, $Prox_{f^*}$ and $Prox_G$ are proximal operators for function $F^*$ and $G$, exact formula of these two operators can be found in \cite{heide2013high}. Once $f$ has been determined, the relative shift between wide-field and confocal image can be accurately calculated by the shift of maximal intensity point of PSF $f$. Accurate image registration can substantially enhance the reconstruction quality.\\

\noindent \textbf{Sample preparation of Immuno-fluorescent staining mouse brain slices.} 
We demonstrated the performance of WFCON-Net on $40\mu m$-thick C57/B6 mouse brain slices obtained from Prof. Yichang Jia's lab, Tsinghua University. The neuron body and microglia of the brain slice are immune-fluorescent stained and the procedures are described as followed.

First, the brain slices were freshly obtained from Leica vibrating microtome 7000 (after perfusion-fixed with $4\%$ paraformaldehyde in 1X PBS), and then incubated with permeate-blocking buffer ($0.3\%$ Triton-X100, $3\%$BSA in 1X PBS) at room temperature for 2 hours. 
After that, the slices were gently washed 5 times (5 min per wash) with washing buffer (0.05\%Tween-20, 3\%BSA in 1X PBS), incubated with NeuN antibody (Cell Signaling \#94403, 100X dilution in washing buffer) and Iba1 antibody (Cell Signaling \#17198, 100X dilution in washing buffer) at 4\textcelsius for 24 hours, protected from light. 
On the next day, the slices were gently washed 5 times again with washing buffer, then incubated with Alexafluor-488 labeled goat-anti-mouse antibody (Cell Signaling \#4408, 200X dilution in washing buffer) and Alexafluor-555 labeled goat-anti-rabbit antibody (Cell Signaling \#4413, 200X dilution in washing buffer) at 4\textcelsius overnight in the dark. On the third day, the slices were gently washed 5 more times with washing buffer, transferred on Superfrost™ Plus slides, mounted with 22 mm No.1.5 square coverslips, and ProLong Gold antifade mountant containing 2 \textmu g/mL DAPI. These prepared slices were protected from light and stored at 4\textcelsius before imaging. All the reagents, coverslips, and tissue slides were purchased from ThermoFisher if mentioned otherwise. \\

\section{Results}
\label{sec:resutls}
\begin{figure*}
\begin{center}
\includegraphics[width=1.0\linewidth]{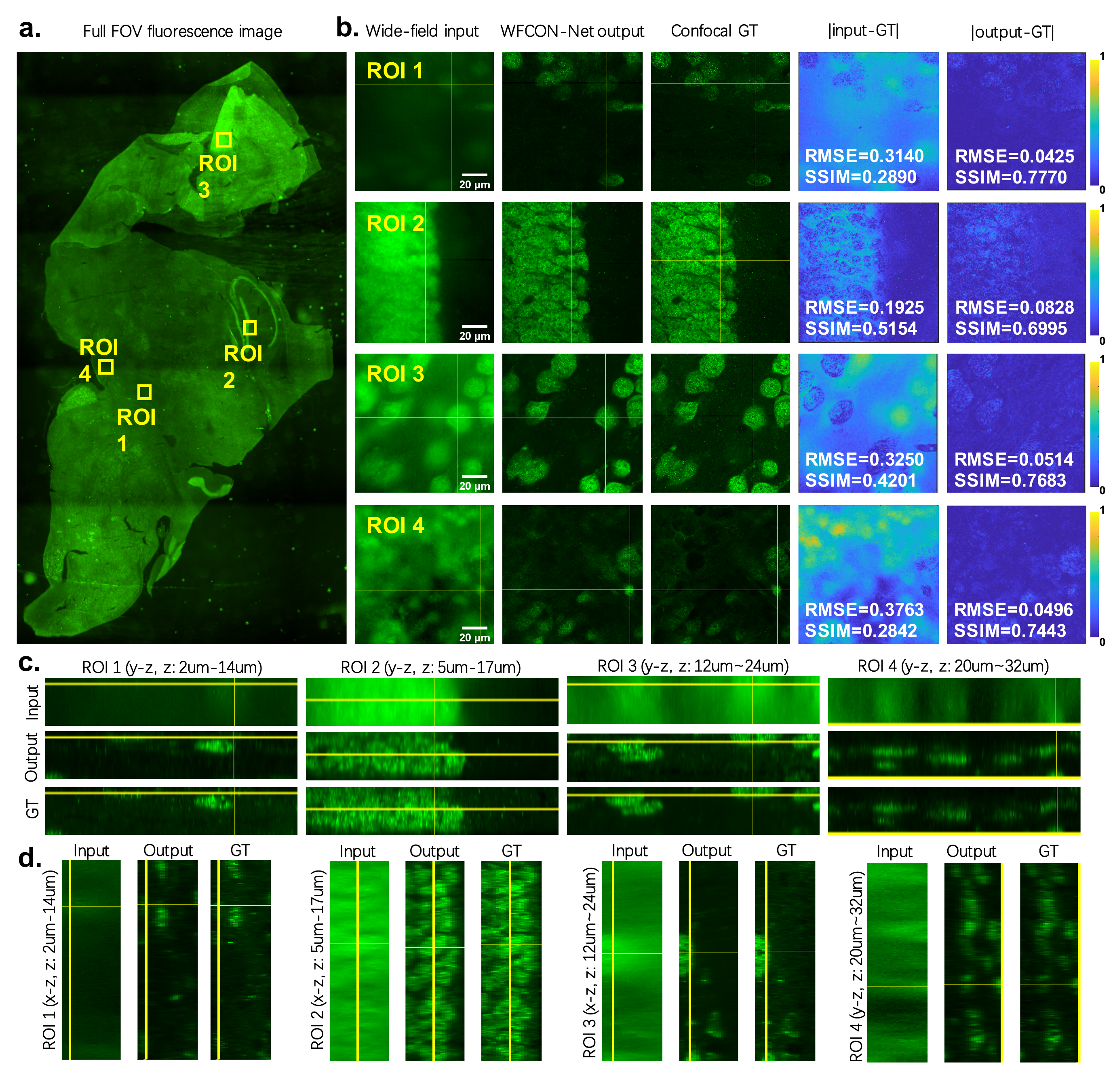}
\end{center}
\caption{\textbf{3D confocal imaging of mouse brain slices using WFCON-Net.}
\textbf{a)}. The full field of view (FOV) axial fluorescence image of a mouse brain slice. The full image is stitched together from 15 sub-images (for different parts of the slice) captured using a 4x/0.2 objective lens.
\textbf{b)}. Digital confocal predictions using WFCON-Net for 4 different ROIs with different neuron densities/structures. The wide-field inputs and confocal ground truths are also shown for comparison. The predicted confocal images match the ground truths (scanned using a confocal microscope) very well at corresponding x-y planes. 
 The root mean square error (RMSE, the lower the better) and the structural similarity index measure (SSIM, the higher the better) are used here to quantitatively evaluate the prediction accuracy. The cross-sectional images span $12\mu m$ in the z-direction (with a step size of $0.5\mu m$).The error maps |output-GT| show that the reconstruction qualities are slightly degraded with the increased neuron aggregation.
 \textbf{c)} Magnified view of y-z cross sections in 4 ROIs same to \textbf{b} show accurate three-dimensional confocal image reconstruction. The z-axis spans of those images are explicitly labeled.
  \textbf{d)} Magnified view of x-z cross sections in 4 ROIs same to \textbf{b}.
 }
\label{fig:mice_brain}
\end{figure*}

\noindent \textbf{3D confocal imaging of mouse brain slices using WFCON-Net.}
We first demonstrate our method on a mouse brain slice, as shown in Figure \ref{fig:mice_brain}a. The model is trained on 31 pairs of the registered wide-field and confocal z-stack images and tested on the other 8 pairs. 
The prediction results, as well as the wide-field inputs, the corresponding confocal ground truths, and the difference images of the selected region of interest (ROI) with different neuron densities/structures are shown in Figure \ref{fig:mice_brain}b. We make use of the root mean square error (RMSE, the lower the better) and the structural similarity index measure (SSIM, the higher the better) to quantitatively evaluate the prediction accuracy. The average RMSE and SSIM of the testing datasets are 0.0575 and 0.7673.
As we can see from the figure, the mice brain sample we applied is a thick scattering sample ($\sim40$ micron), that the details of the wide-field images are completely overwhelmed by the scattering background. The predictions of such highly-scattering samples are significantly challenging than thinner samples ($\sim$ several microns) \cite{wu2019three,ning2020deep}. Our proposed GAN-based WFCON-Net, with 3D convolutional operations, can successfully reconstruct the high-contrast, high-resolution z-stack images from the wide-field captures, matching the confocal images well at the corresponding planes. The magnified y-z, x-z cross-sections of the image stacks (span $12\mu m$ in the z-direction, the full-stack results (over $38\mu m$) are shown in video 1 and 2 in the supplementary), in Figure \ref{fig:mice_brain}c and d, demonstrate the reconstruction accuracy across z-axis. The performance for the areas with denser neuron accumulation (such as ROI2) degrades slightly because of the more rigorous scattering background, leading to a larger reconstruction RMSE and a lower SSIM.

\begin{figure*}
\begin{center}
\includegraphics[width=1.0\linewidth]{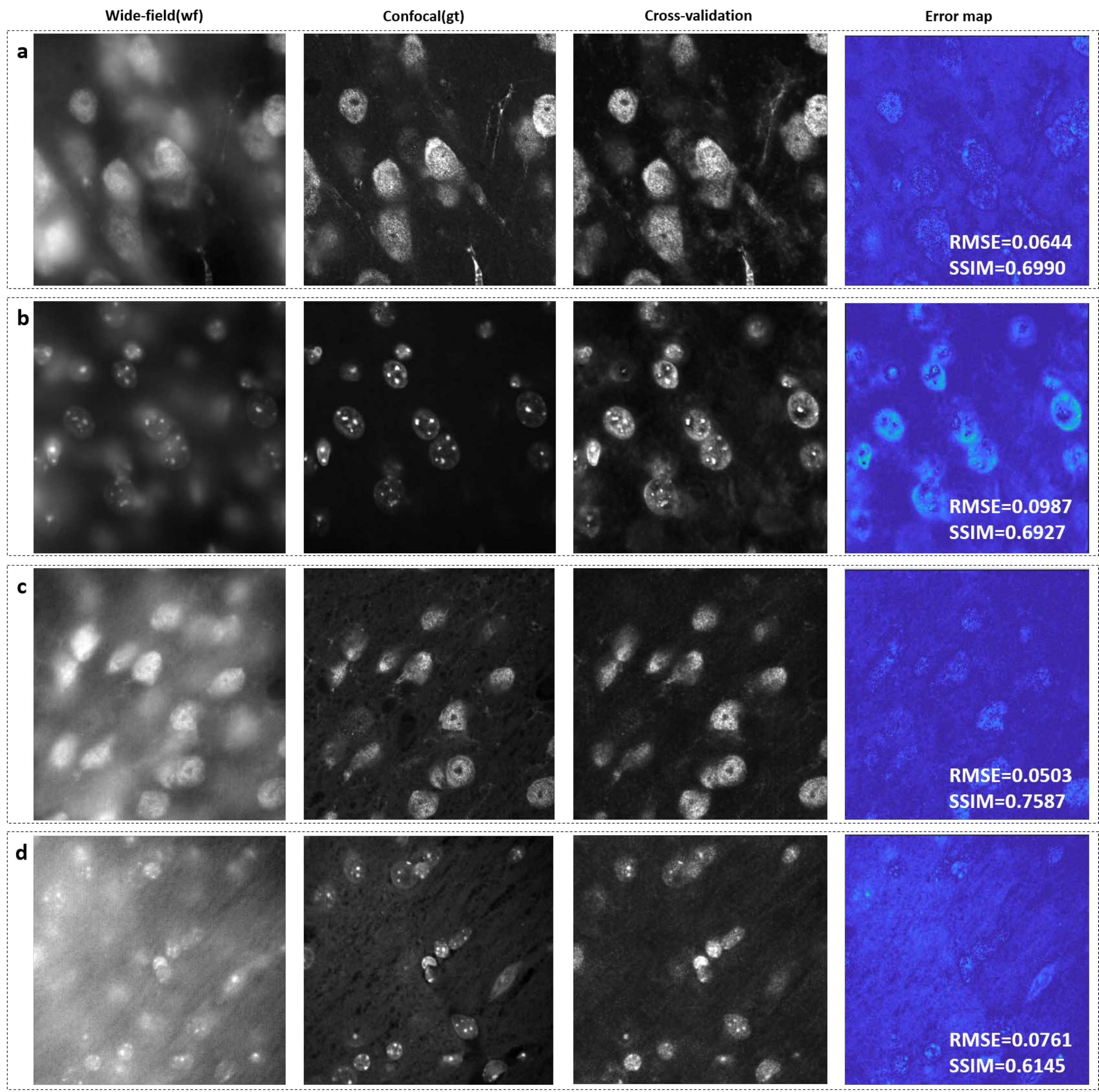}
\end{center}
\caption{
\textbf{Ablation study on model generalization ability.}
a,b)The green channel(GFP) and blue channel(DAPI) of wide-field fluorescence images are captured from another neuron slice with less background. The results show good background noise suppression capability with acceptable artifacts. c,d) The same setting but using another slice with more scattering and background than training data. As these images are more similar to the training data, their inference results are better than those images with less background.
   }
\label{fig:sample_general}
\end{figure*}

Our network also shows good generalization ability to unseen data. Without retraining, we tested model with images captured from another two neuron slices, which have obviously different levels of background and scattering. Images of one slice have less background (fig. \ref{fig:sample_general}a,b)) while another suffer from more severe background (fig. \ref{fig:sample_general}c,d)). The images are shown in grayscale to clarify the level of background noise. The reconstructed images show good background suppression capability with acceptable artifacts. Images of green channel(fig. \ref{fig:sample_general}a,c), GFP) exhibit higher accuracy than blue channel(fig. \ref{fig:sample_general}b,d), DAPI) as the same channel used for the training. As images with more severe background share more similarities with the training set, their inference results are better than those images with less background. The average RMSE and SSIM of the testing datasets are 0.0695/0.7153 and 0.0544/0.7501 for green channel and 0.1106/0.6712 and 0.0923/0.6527 for blue channel.

\begin{figure*}
\begin{center}
\includegraphics[width=1.0\linewidth]{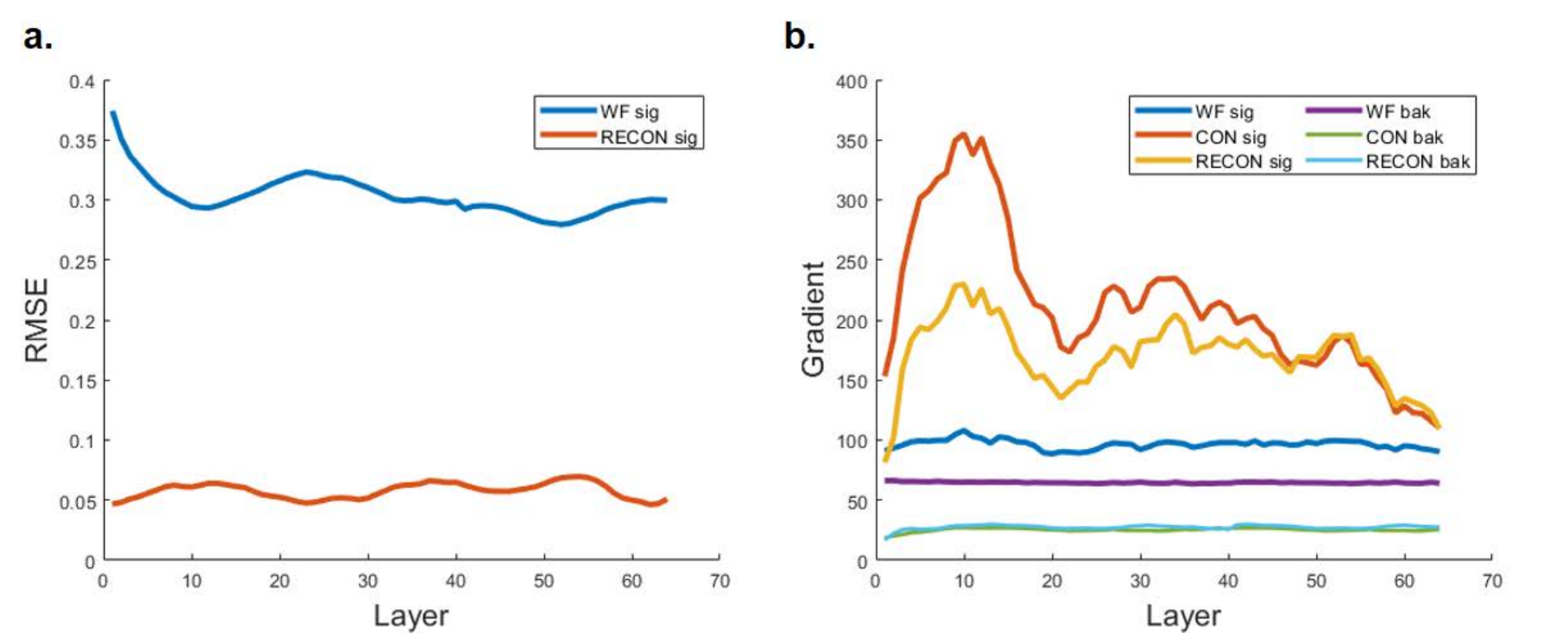}
\end{center}
\caption{
\textbf{Quantitative analysis of network reconstruction along with depth}
a)The average RMSE of wide-field images and recovered images refering to confocal ground-truth images along with depth. b)The average sharpness curve of neuron slice wide-field images, confocal images, reconstruction images and the corresponding backgrounds.}
\label{fig:depth}
\end{figure*}

The image quality of inputs together with outputs are affected by the thickness of sample. As axial depth increases, the sharpness of confocal ground-truth images and the reconstructions are both deteriorated. However, as shown in Figure \ref{fig:depth}a, the learning to the confocal ground-truth remains stable with different depth, which confirms the learning robustness of the proposed network. Furthermore, we measure the sharpness by calculating the gradient of image patches of size 32x32 by $\Delta I=|\Delta I_x|+|\Delta I_y|$. In each depth, we take the patch with largest gradient as a sharpness measurement of signal, and the smallest one as a measurement of background noise. As shown in Figure \ref{fig:depth}b, the sharpness of the background keep almost constant while the sharpness of confocal images is degraded along with depth. Although in superficial layers the confocal images outperform the network outputs, but they continuously lost superiority when image deeper because the network can take advantage of a priori knowledge learned from shallower layer, which is a great advantage of our Deep learning enabled confocal microscope.\\

\begin{figure*}
\begin{center}
\includegraphics[width=1.0\linewidth]{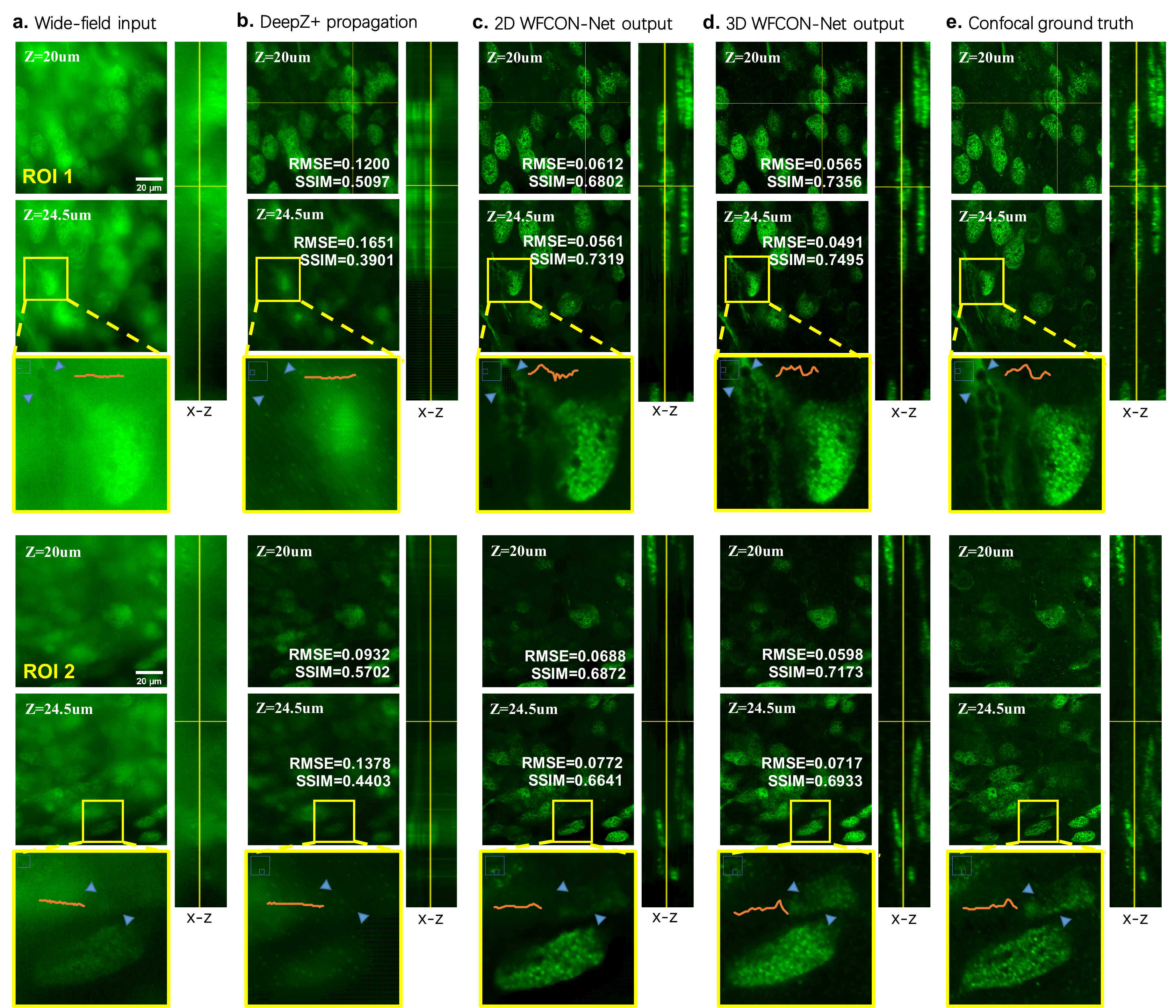}
\end{center}
\caption{
\textbf{Comparison with other wide-field to confocal cross-modality methods.}
\textbf{a).} Wide-field input images of two ROIs. 
\textbf{b).} Propagated 3D confocal images using DeepZ+ \cite{wu2019three}. The DeepZ+ takes the single-layer wide-field image (eg. layer $z=20\mu m$) as input, and outputs the propagated 3D confocal z-stacks (only layer $z=24.5\mu m$ is shown). The propagation fails to generate accurate predictions for thick samples (due to the distributed fluorescence signals in multiple layers and the strong scattering background).
\textbf{C).} Predicted confocal images using WFCON-Net2D. The WFCON-Net2D (with 2D Conv blocks) takes the wide-field z-stacks as input, and predicts the corresponding confocal images in a layer-to-layer manner.
\textbf{d).} Predicted confocal images using WFCON-Net. 
The WFCON-Net (with 3D Conv blocks) benefits from the stronger representation ability and inter-layer correlation information, and thus surpasses 2D methods in confocal predictions with higher accuracy and richer details, which can be verified by the line profile marked by two triangular arrows in the insets of images.
\textbf{e).} Confocal ground truth images. 
For all the images, the x-y images at $z=20\mu m, 24.5\mu m$ and their corresponding x-z cross-sections are shown.
}
\label{fig:compare}
\end{figure*}

\noindent \textbf{WFCON-Net outperforms 2D convolutional networks for thick-sample prediction.}
Next, we compare our method with other 2D convolutional GAN-based networks. The prediction results for DeepZ+ \cite{wu2019three}, the 2D version of WFCON-Net and our 3D WFCON-Net are shown in Figure \ref{fig:compare}. 
We implemented the DeepZ+ propagation algorithm as proposed in \cite{wu2019three}, which takes the single-layer wide-field image as input. The algorithm propagates the single wide-field image to predict multi-layer confocal z-stacks. Such propagation works well on the thin/sparse BPAEC microtubule structures, but fails to generate accurate propagation for thick/dense mice brain samples as shown in Figure \ref{fig:compare}b. 
We also investigated the 2D WFCON-Net by replacing the 3D convolutional blocks with 2D convolutions. Figure \ref{fig:compare}c and d display the predicted confocal images of 2D WFCON-Net and 3D WFCON-Net. The yellow boxes show a magnified view of the cell body and the insets mark the intensity profile of specific regions. 
Our 3D convolutional WFCON-Net benefits from the stronger representation ability and inter-layer correlation information, outperforming the 2D methods in confocal predictions with less blur, richer details and higher accuracy.\\

\begin{figure*}
\begin{center}
\includegraphics[width=1.0\linewidth]{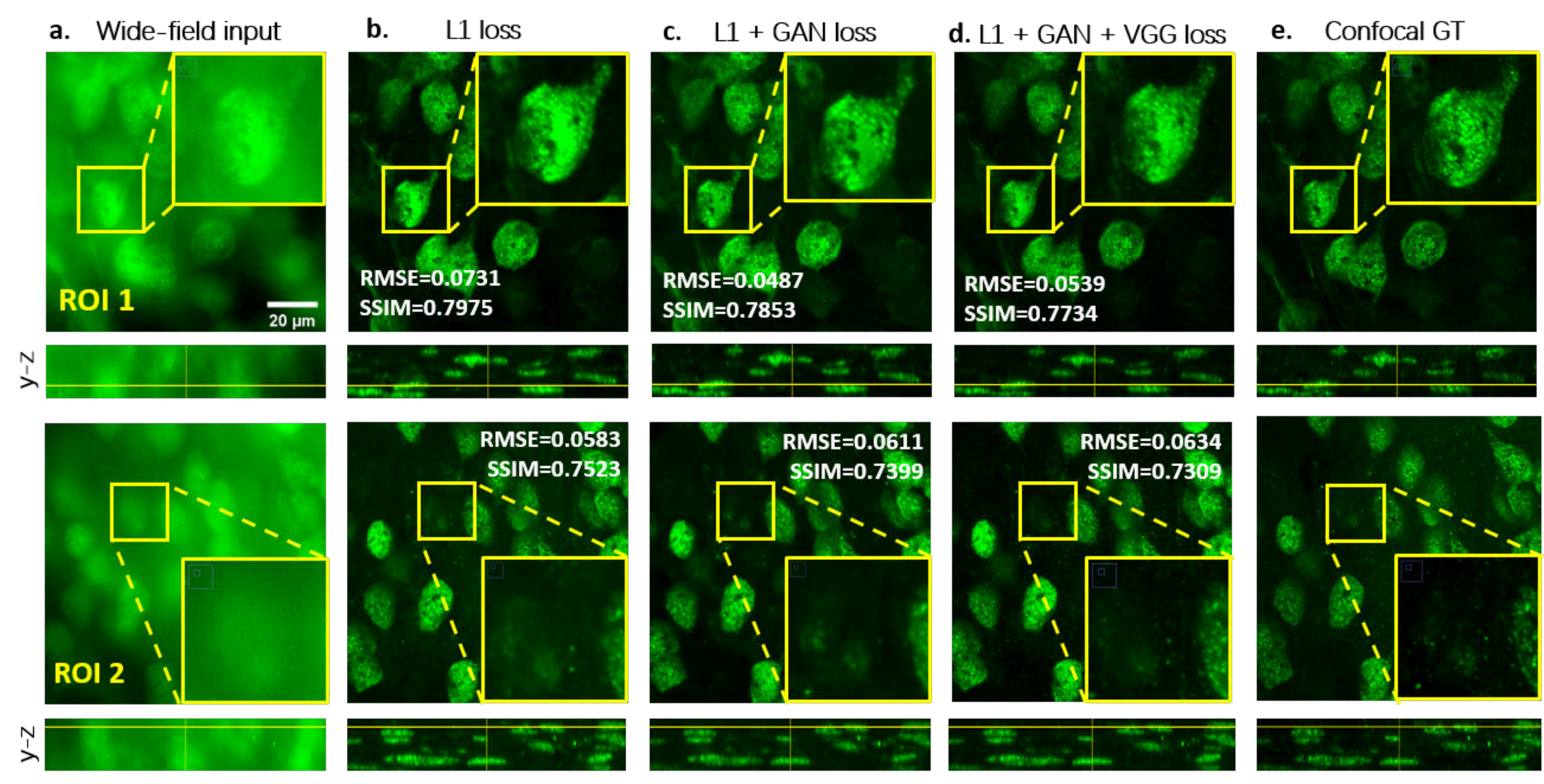}
\end{center}
\caption{\textbf{Training with GAN and VGG loss.}
We use the sum of L1 loss, VGG loss and GAN loss as our loss function during training. The use of perceptual loss (VGG loss) encourages predictions with higher quality and richer details.
\textbf{a).} Wide-field input inputs of two different ROIs.
\textbf{b).} Predicted confocal images of the network trained with only L1 loss\cite{ning2020deep}.
\textbf{c).} Predicted confocal images of the network trained with L1 loss and GAN loss\cite{wu2019three}.
\textbf{d).} Our method, trained with a combination of L1 loss, VGG loss and GAN loss.
\textbf{e).} Confocal ground truth images. Although the use of non-PSNR oriented perceptual loss will slightly affect SSIMs and RMSEs, but as shown in the magnified views, our method can reconstruct more details (especially in the overexposed areas -- ROI1) and suppress more background noises (ROI2). 
}
\label{fig:vggloss}
\end{figure*}

\noindent \textbf{Training with GAN and VGG loss.}
Moreover, we introduce the perceptual VGG loss together with GAN loss to train our network. VGG loss is prevalent in natural image super-resolution as it can enhance the high frequency details that are filtered in low-resolution images, making the image more realistic\cite{2016Photo}. On the contrary, conventional PSNR oriented loss tends to smooth the reconstruction result\cite{johnson2016perceptual}. As shown in Figure \ref{fig:vggloss}, by adding the VGG loss, the reconstruction details are well preserved (especially in the overexposed area) and the background noises are well suppressed. The use of the perceptual loss encourages high-quality, high-resolution predictions, matching the corresponding confocal images well. The result also demonstrates that different image generation tasks share some feature similarities.\\

\begin{figure*}
\begin{center}
\includegraphics[width=1.0\linewidth]{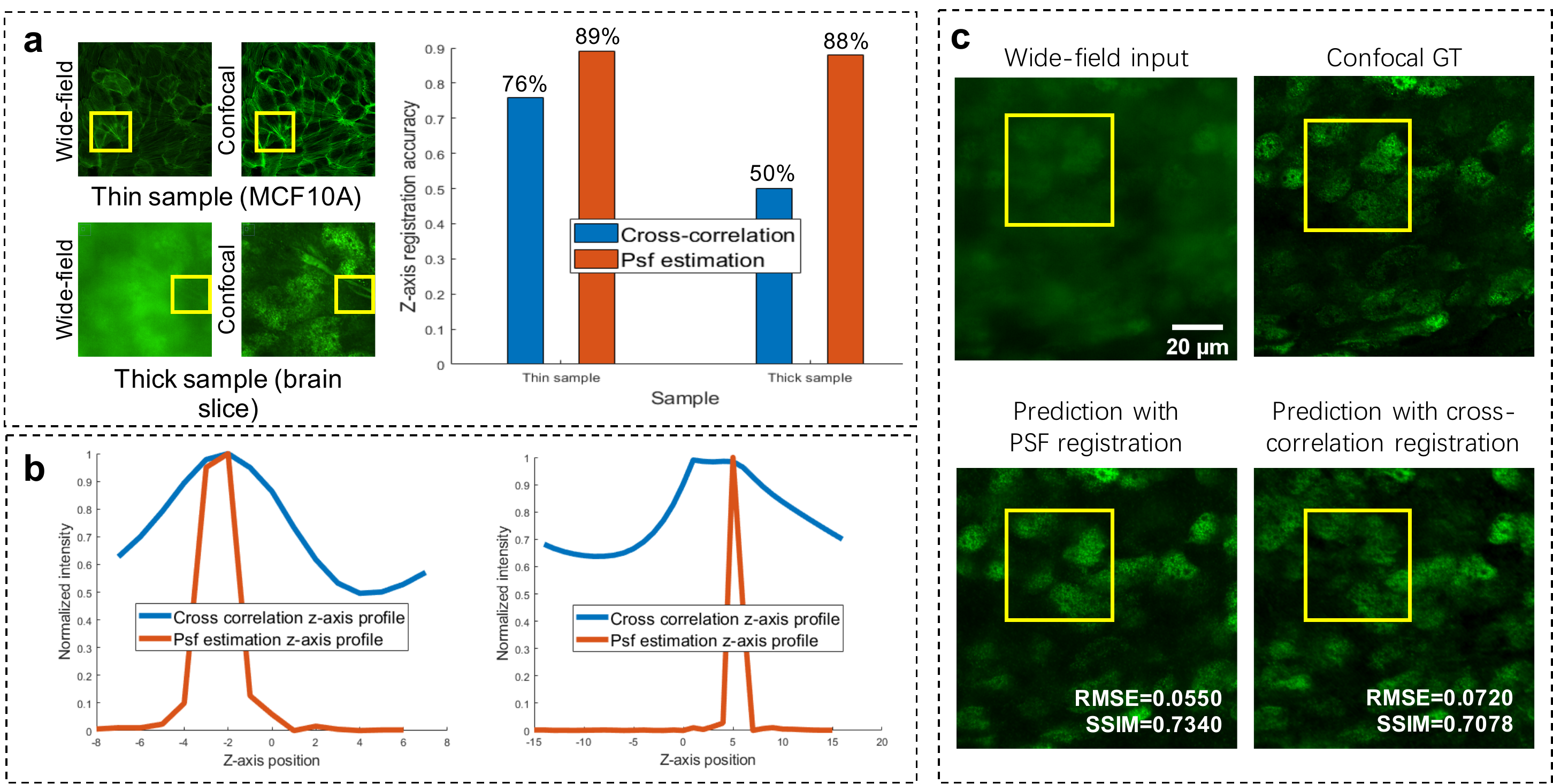}
\end{center}
\caption{
\textbf{Image registration with estimated PSF.}
We register images by estimating the 3D point spread functions (PSF) from paired wide-field and confocal z-stack images. The peak of the calculated PSF profile is then used to determine the lateral and axial shifts between unregistered images. 
\textbf{a).} We compare the PSF registration method with the cross-correlation method for thin sample (MCF10A, 3-microns thickness) and thick sample (brain slice, 38-microns thickness), respectively. The PSF registration (orange) outperforms the cross-correlation method (blue) in higher z-axis accuracy, especially for the thick samples.
\textbf{b).} Profiles along the z-axis of cross-correlation and calculated PSF for the thin sample (left) and the thick sample (right). The sharp peak of PSF benefits the accurate registration and is robust to noises across wide-field and confocal images.
 \textbf{c).} WFCON-Net predictions with and without PSF registered training data. The PSF registration method improves the reconstruction quality. 
The ground-truth registrations are manually aligned.
}
\label{fig:registration}
\end{figure*}

\noindent \textbf{PSF-based registrations.}
\noindent Image registration is one of the main issues for learning-based cross-modality image reconstruction. The greater the degree of scattering and noise, the higher the difficulty for registration. As shown in (fig\ref{fig:registration}.a)), fluorescent structures in the thick sample are more ambiguous than the thin sample, resulting in the decrease of the accuracy of registration and the degradation of the reconstruction performance. We can observe that the cross-correlation curve along the z-axis is flatter than the PSF curve (fig\ref{fig:registration}.b), especially in thick samples. This flatness causes difficulty identifying the actual shift between image pairs. By calculating the PSF between 3D wide-field and confocal image pairs, we find the physical connection between two modalities and the registration process becomes more accurate, and reconstructed details are thus well preserved (fig\ref{fig:registration}.c). The registration error of the PSF estimation method occurs only when the shift between wide-field and confocal images falls in the middle of two integer numbers, corresponding to two peaks with approximate height as shown in the left plot of (fig\ref{fig:registration}.b). Hence, the maximum theoretical error is nearly half of the minimum z-step size. The PSF registration error can be further reduced with a smaller z-step size. \\

\begin{figure*}
\begin{center}
\includegraphics[width=1.0\linewidth]{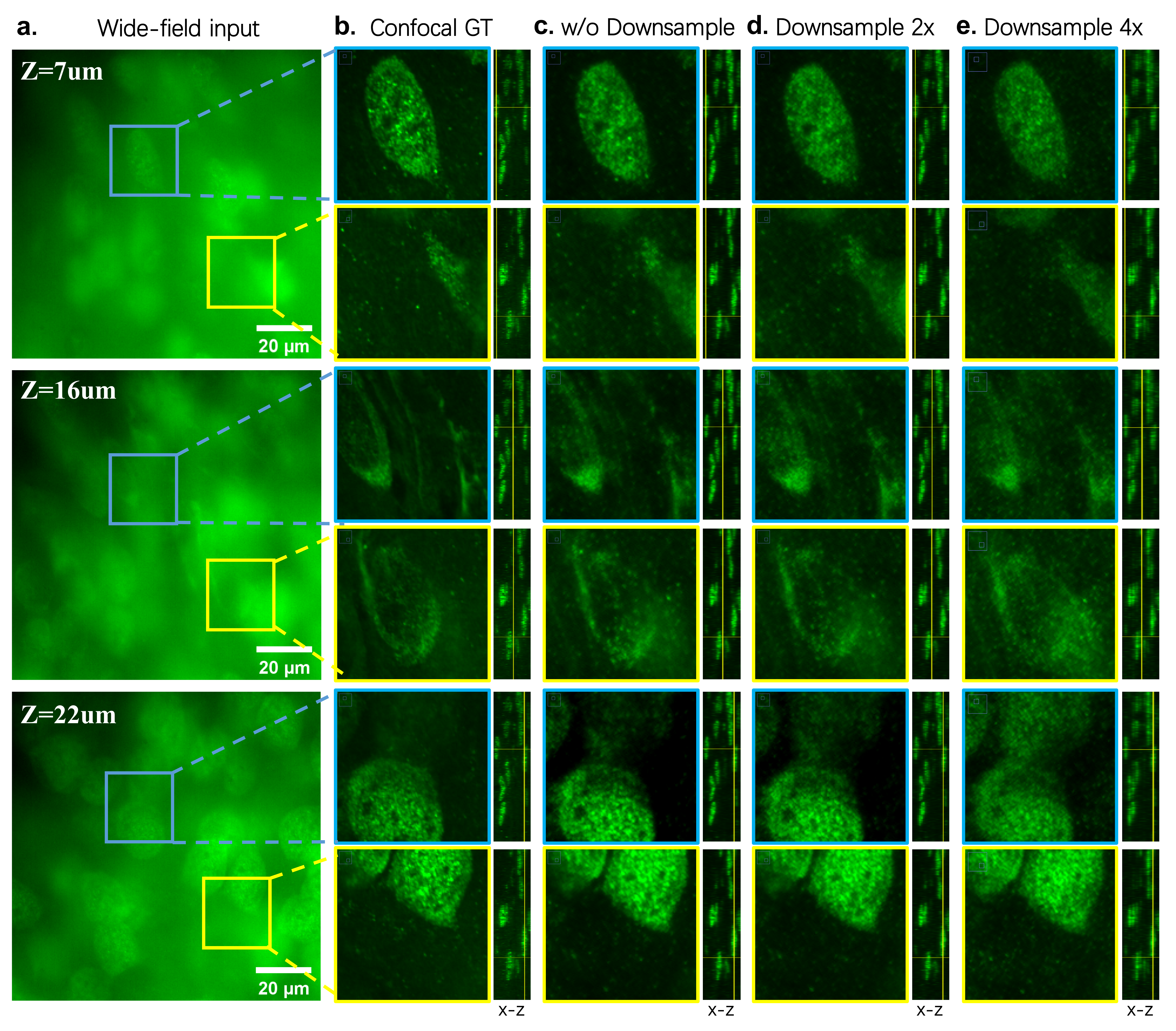}
\end{center}
\caption{
\textbf{Ablation study on the number of wide-field input layers (along z).}
We test our algorithm with the different levels of input layers by downsampling the wide-field z-stacks, without retraining the network. The whole z-stacks are reinterpolated from the downsampled data before input to our network. 
\textbf{a) } Wide-field inputs at different depths. Two specific regions of interest (bounded by the blue and yellow boxes) are enlarged to show details.
The x-z cross-sections are also shown for localization.
\textbf{b) } Confocal (ground truth, GT) images of the two enlarged regions.
\textbf{c) } WFCON-Net reconstructed images without downsampling.
\textbf{d) } WFCON-Net reconstructed images with 2x downsampled input and interpolation. The results are comparable to the results without downsampling.
\textbf{e) } WFCON-Net reconstructed images with 4x downsampled input and interpolation. The reconstruction accuracy degrades slightly and mainly in out-of-focus layers, since these layers are more vulnerable to the background signals originated from the in-focus layers.
}
\label{fig:interpolation}
\end{figure*}

\noindent \textbf{Ablation study on the number of wide-field input layers.}
Our method requires capturing the whole stack of wide-field images by scanning along the z-direction. 
To accommodate the applications that demand faster data acquisition, we investigated the performance of our network trained with whole dataset but tested with fewer images as input. 
Specifically, we subsample the layers of captured z-stacks at an interval of 2 layers (downsample 2x) and 4 layers (downsample 4x), and interpolated the missing layers to the original size before feeding them into the network. The downsampled z-stacks come with an equivalent z-step of $1 \mu m$ and $2 \mu m$, respectively (original z-step $\sim 0.5\mu m$).
Figure \ref{fig:interpolation} shows the reconstruction results with 2x downsampling, 4x downsampling, and without downsampling (without retraining the network). 
We can see that the model trained for the original data (without downsampling) works well for the 2x downsampling, and is slightly degraded for 4x downsampled z-stacks. 
The spacing of 4x downsampled images in z-axis is comparable to the spacing Huang et al. demonstrated \cite{huang2021recurrent}, but we show good results on samples with much more complicated structures and unwanted backgrounds.
The degradation is mainly caused by the gap between captured input layers and the interpolated layers. 
Higher reconstruction quality can be obtained if we retrain the network with the interpolation process included in the pipeline. 
The average RMSE and SSIM of the testing dataset for without/2x/4x downsampling are 0.0575/0.0598/0.0657 and 0.7673/0.7343/0.6652, respectively.\\

\section{Discussion}
\label{sec:discussion}
We provide a GAN-based 3D improved U-Net neural network to generate the confocal images from the wide-field images. To the best of our knowledge, we are the first to deal with wide-field images of a thick and complex sample, in which input image quality is severely degraded by scattering and background noise. 
By using GAN-based 3D U-Net with additional residual mapping, normalization layer and VGG loss, together with accurate image registration, the reconstructed details are repaired from wide-field images. 
Our model has shown good generalization ability, either channel generalization or sample generalization, and can also be used with z-downsampled inputs. 
Our experimental results demonstrate generalization ability to handle unseen data, stability in the reconstruction results, high spatial resolution even when imaging thick (~40 microns) highly-scattering samples.
We believe that such learning-based-microscopes have the potential to democratize scientific imaging bringing confocal quality imaging to every lab that has a wide-field microscope.

Our method is currently limited by the requirement of the training dataset to achieve the best performance. 
In some scenarios, like in-vivo neuron activity imaging of moving mice, the ground truth confocal or two-photon images are hard to acquire and register to the wide-field inputs. 
If we using the model trained on another sample, the inference result will be degraded. To deal with this problem, we can generate images from simulations as in\cite{2018Efficient} or explore variants of transfer learning/domain adaptation.

\section*{Acknowledgement}
We would like to thank Yichang Jia and Xu Zhang from Tsinghua University for their help on preparing the samples. We would also like to thank
Xu Chen from Tsinghua Shenzhen International Graduate School for preparing the visualization video. In addition, we thank Yanli Zhang and Yalan Chen from Technology Center for Protein Sciences of Tsinghua University for assistance of using microscopes.

\section*{Disclosures}
The authors declare no conflicts of interest.

\bibliography{references}

\begin{thebibliography}{10}
\newcommand{\enquote}[1]{``#1''}

\bibitem{wilson1990confocal}
T.~Wilson, \enquote{Confocal microscopy,} {\protect\JournalTitle{San Diego}}
  (1990).

\bibitem{pawley2006handbook}
J.~Pawley, \emph{Handbook of biological confocal microscopy}, vol. 236
  (Springer Science \& Business Media, 2006).

\bibitem{helmchen2005deep}
F.~Helmchen and W.~Denk, \enquote{Deep tissue two-photon microscopy,}
  {\protect\JournalTitle{Nature methods}} \textbf{2}, 932--940 (2005).

\bibitem{so2000two}
P.~T. So, C.~Y. Dong, B.~R. Masters, and K.~M. Berland, \enquote{Two-photon
  excitation fluorescence microscopy,} {\protect\JournalTitle{Annual review of
  biomedical engineering}} \textbf{2}, 399--429 (2000).

\bibitem{park2015high}
J.-H. Park, W.~Sun, and M.~Cui, \enquote{High-resolution in vivo imaging of
  mouse brain through the intact skull,} {\protect\JournalTitle{Proceedings of
  the National Academy of Sciences}} \textbf{112}, 9236--9241 (2015).

\bibitem{lu2017video}
R.~Lu, W.~Sun, Y.~Liang, A.~Kerlin, J.~Bierfeld, J.~D. Seelig, D.~E. Wilson,
  B.~Scholl, B.~Mohar, M.~Tanimoto \emph{et~al.}, \enquote{Video-rate
  volumetric functional imaging of the brain at synaptic resolution,}
  {\protect\JournalTitle{Nature neuroscience}} \textbf{20}, 620--628 (2017).

\bibitem{papadopoulos2020dynamic}
I.~N. Papadopoulos, J.-S. Jouhanneau, N.~Takahashi, D.~Kaplan, M.~Larkum,
  J.~Poulet, and B.~Judkewitz, \enquote{Dynamic conjugate f-sharp microscopy,}
  {\protect\JournalTitle{Light: Science \& Applications}} \textbf{9}, 1--8
  (2020).

\bibitem{zong2021miniature}
W.~Zong, R.~Wu, S.~Chen, J.~Wu, H.~Wang, Z.~Zhao, G.~Chen, R.~Tu, D.~Wu, Y.~Hu
  \emph{et~al.}, \enquote{Miniature two-photon microscopy for enlarged
  field-of-view, multi-plane and long-term brain imaging,}
  {\protect\JournalTitle{Nature Methods}} \textbf{18}, 46--49 (2021).

\bibitem{chen2014lattice}
B.-C. Chen, W.~R. Legant, K.~Wang, L.~Shao, D.~E. Milkie, M.~W. Davidson,
  C.~Janetopoulos, X.~S. Wu, J.~A. Hammer, Z.~Liu \emph{et~al.},
  \enquote{Lattice light-sheet microscopy: imaging molecules to embryos at high
  spatiotemporal resolution,} {\protect\JournalTitle{Science}} \textbf{346}
  (2014).

\bibitem{keller2008reconstruction}
P.~J. Keller, A.~D. Schmidt, J.~Wittbrodt, and E.~H. Stelzer,
  \enquote{Reconstruction of zebrafish early embryonic development by scanned
  light sheet microscopy,} {\protect\JournalTitle{science}} \textbf{322},
  1065--1069 (2008).

\bibitem{power2017guide}
R.~M. Power and J.~Huisken, \enquote{A guide to light-sheet fluorescence
  microscopy for multiscale imaging,} {\protect\JournalTitle{Nature methods}}
  \textbf{14}, 360--373 (2017).

\bibitem{gao2012noninvasive}
L.~Gao, L.~Shao, C.~D. Higgins, J.~S. Poulton, M.~Peifer, M.~W. Davidson,
  X.~Wu, B.~Goldstein, and E.~Betzig, \enquote{Noninvasive imaging beyond the
  diffraction limit of 3d dynamics in thickly fluorescent specimens,}
  {\protect\JournalTitle{Cell}} \textbf{151}, 1370--1385 (2012).

\bibitem{mcdole2018toto}
K.~McDole, L.~Guignard, F.~Amat, A.~Berger, G.~Malandain, L.~A. Royer, S.~C.
  Turaga, K.~Branson, and P.~J. Keller, \enquote{In toto imaging and
  reconstruction of post-implantation mouse development at the single-cell
  level,} {\protect\JournalTitle{Cell}} \textbf{175}, 859--876 (2018).

\bibitem{yue2020long}
Y.~Yue, W.~Zong, X.~Li, J.~Li, Y.~Zhang, R.~Wu, Y.~Liu, J.~Cui, Q.~Wang,
  Y.~Bian \emph{et~al.}, \enquote{Long-term, in toto live imaging of
  cardiomyocyte behaviour during mouse ventricle chamber formation at
  single-cell resolution,} Tech. rep., Nature Publishing Group (2020).

\bibitem{rivenson2017deep}
Y.~Rivenson, Z.~G{\"o}r{\"o}cs, H.~G{\"u}naydin, Y.~Zhang, H.~Wang, and
  A.~Ozcan, \enquote{Deep learning microscopy,} {\protect\JournalTitle{Optica}}
  \textbf{4}, 1437--1443 (2017).

\bibitem{wang2019deep}
H.~Wang, Y.~Rivenson, Y.~Jin, Z.~Wei, R.~Gao, H.~G{\"u}nayd{\i}n, L.~A.
  Bentolila, C.~Kural, and A.~Ozcan, \enquote{Deep learning enables
  cross-modality super-resolution in fluorescence microscopy,}
  {\protect\JournalTitle{Nature methods}} \textbf{16}, 103--110 (2019).

\bibitem{zhang2020exceeding}
H.~Zhang, Y.~Zhao, C.~Fang, G.~Li, M.~Zhang, Y.-H. Zhang, and P.~Fei,
  \enquote{Exceeding the limits of 3d fluorescence microscopy using a
  dual-stage-processing network,} {\protect\JournalTitle{Optica}} \textbf{7},
  1627--1640 (2020).

\bibitem{goy2018low}
A.~Goy, K.~Arthur, S.~Li, and G.~Barbastathis, \enquote{Low photon count phase
  retrieval using deep learning,} {\protect\JournalTitle{Physical review
  letters}} \textbf{121}, 243902 (2018).

\bibitem{weigert2018content}
M.~Weigert, U.~Schmidt, T.~Boothe, A.~M{\"u}ller, A.~Dibrov, A.~Jain,
  B.~Wilhelm, D.~Schmidt, C.~Broaddus, S.~Culley \emph{et~al.},
  \enquote{Content-aware image restoration: pushing the limits of fluorescence
  microscopy,} {\protect\JournalTitle{Nature methods}} \textbf{15}, 1090--1097
  (2018).

\bibitem{ouyang2018deep}
W.~Ouyang, A.~Aristov, M.~Lelek, X.~Hao, and C.~Zimmer, \enquote{Deep learning
  massively accelerates super-resolution localization microscopy,}
  {\protect\JournalTitle{Nature biotechnology}} \textbf{36}, 460--468 (2018).

\bibitem{nehme2020deepstorm3d}
E.~Nehme, D.~Freedman, R.~Gordon, B.~Ferdman, L.~E. Weiss, O.~Alalouf, T.~Naor,
  R.~Orange, T.~Michaeli, and Y.~Shechtman, \enquote{Deepstorm3d: dense 3d
  localization microscopy and psf design by deep learning,}
  {\protect\JournalTitle{Nature Methods}} \textbf{17}, 734--740 (2020).

\bibitem{li2020fast}
X.~Li, J.~Dong, B.~Li, Y.~Zhang, Y.~Zhang, A.~Veeraraghavan, and X.~Ji,
  \enquote{Fast confocal microscopy imaging based on deep learning,} in
  \emph{2020 IEEE International Conference on Computational Photography
  (ICCP),}  (IEEE, 2020), pp. 1--12.

\bibitem{wu2018extended}
Y.~Wu, Y.~Rivenson, Y.~Zhang, Z.~Wei, H.~G{\"u}naydin, X.~Lin, and A.~Ozcan,
  \enquote{Extended depth-of-field in holographic imaging using
  deep-learning-based autofocusing and phase recovery,}
  {\protect\JournalTitle{Optica}} \textbf{5}, 704--710 (2018).

\bibitem{pinkard2019deep}
H.~Pinkard, Z.~Phillips, A.~Babakhani, D.~A. Fletcher, and L.~Waller,
  \enquote{Deep learning for single-shot autofocus microscopy,}
  {\protect\JournalTitle{Optica}} \textbf{6}, 794--797 (2019).

\bibitem{ounkomol2018label}
C.~Ounkomol, S.~Seshamani, M.~M. Maleckar, F.~Collman, and G.~R. Johnson,
  \enquote{Label-free prediction of three-dimensional fluorescence images from
  transmitted-light microscopy,} {\protect\JournalTitle{Nature methods}}
  \textbf{15}, 917--920 (2018).

\bibitem{christiansen2018silico}
E.~M. Christiansen, S.~J. Yang, D.~M. Ando, A.~Javaherian, G.~Skibinski,
  S.~Lipnick, E.~Mount, A.~O’Neil, K.~Shah, A.~K. Lee \emph{et~al.},
  \enquote{In silico labeling: predicting fluorescent labels in unlabeled
  images,} {\protect\JournalTitle{Cell}} \textbf{173}, 792--803 (2018).

\bibitem{rivenson2019virtual}
Y.~Rivenson, H.~Wang, Z.~Wei, K.~de~Haan, Y.~Zhang, Y.~Wu, H.~G{\"u}nayd{\i}n,
  J.~E. Zuckerman, T.~Chong, A.~E. Sisk \emph{et~al.}, \enquote{Virtual
  histological staining of unlabelled tissue-autofluorescence images via deep
  learning,} {\protect\JournalTitle{Nature biomedical engineering}} \textbf{3},
  466 (2019).

\bibitem{wu2019three}
Y.~Wu, Y.~Rivenson, H.~Wang, Y.~Luo, E.~Ben-David, L.~A. Bentolila, C.~Pritz,
  and A.~Ozcan, \enquote{Three-dimensional virtual refocusing of fluorescence
  microscopy images using deep learning,} {\protect\JournalTitle{Nature
  methods}} \textbf{16}, 1323--1331 (2019).

\bibitem{Xiaoyu2018Deep}
X.~Zhang, Y.~Chen, K.~Ning, C.~Zhou, Y.~Han, H.~Gong, and J.~Yuan,
  \enquote{Deep learning optical-sectioning method.}
  {\protect\JournalTitle{Optics Express}}  (2018).

\bibitem{9136890}
S.~{Lim}, H.~{Park}, S.~{Lee}, S.~{Chang}, B.~{Sim}, and J.~C. {Ye},
  \enquote{Cyclegan with a blur kernel for deconvolution microscopy: Optimal
  transport geometry,} {\protect\JournalTitle{IEEE Transactions on
  Computational Imaging}} \textbf{6}, 1127--1138 (2020).

\bibitem{ning2020deep}
K.~Ning, X.~Zhang, X.~Gao, T.~Jiang, H.~Wang, S.~Chen, A.~Li, and J.~Yuan,
  \enquote{Deep-learning-based whole-brain imaging at single-neuron
  resolution,} {\protect\JournalTitle{Biomedical Optics Express}} \textbf{11},
  3567--3584 (2020).

\bibitem{lim2020three}
J.~Lim, A.~B. Ayoub, and D.~Psaltis, \enquote{Three-dimensional tomography of
  red blood cells using deep learning,} {\protect\JournalTitle{Advanced
  Photonics}} \textbf{2}, 026001 (2020).

\bibitem{huang2021recurrent}
L.~Huang, H.~Chen, Y.~Luo, Y.~Rivenson, and A.~Ozcan, \enquote{Recurrent neural
  network-based volumetric fluorescence microscopy,}
  {\protect\JournalTitle{Light: Science \& Applications}} \textbf{10}, 1--16
  (2021).

\bibitem{cciccek20163d}
{\"O}.~{\c{C}}i{\c{c}}ek, A.~Abdulkadir, S.~S. Lienkamp, T.~Brox, and
  O.~Ronneberger, \enquote{3d u-net: learning dense volumetric segmentation
  from sparse annotation,} in \emph{International conference on medical image
  computing and computer-assisted intervention,}  (Springer, 2016), pp.
  424--432.

\bibitem{ulyanov2016instance}
D.~Ulyanov, A.~Vedaldi, and V.~Lempitsky, \enquote{Instance normalization: The
  missing ingredient for fast stylization,} {\protect\JournalTitle{arXiv
  preprint arXiv:1607.08022}}  (2016).

\bibitem{glorot2011deep}
X.~Glorot, A.~Bordes, and Y.~Bengio, \enquote{Deep sparse rectifier neural
  networks,} in \emph{Proceedings of the fourteenth international conference on
  artificial intelligence and statistics,}  (2011), pp. 315--323.

\bibitem{mao2017least}
X.~Mao, Q.~Li, H.~Xie, R.~Y. Lau, Z.~Wang, and S.~Paul~Smolley, \enquote{Least
  squares generative adversarial networks,} in \emph{Proceedings of the IEEE
  international conference on computer vision,}  (2017), pp. 2794--2802.

\bibitem{johnson2016perceptual}
J.~Johnson, A.~Alahi, and L.~Fei-Fei, \enquote{Perceptual losses for real-time
  style transfer and super-resolution,} in \emph{European conference on
  computer vision,}  (Springer, 2016), pp. 694--711.

\bibitem{heide2013high}
F.~Heide, M.~Rouf, M.~B. Hullin, B.~Labitzke, W.~Heidrich, and A.~Kolb,
  \enquote{High-quality computational imaging through simple lenses,}
  {\protect\JournalTitle{ACM Transactions on Graphics (TOG)}} \textbf{32},
  1--14 (2013).

\bibitem{chambolle2011first}
A.~Chambolle and T.~Pock, \enquote{A first-order primal-dual algorithm for
  convex problems with applications to imaging,} {\protect\JournalTitle{Journal
  of mathematical imaging and vision}} \textbf{40}, 120--145 (2011).

\bibitem{2016Photo}
C.~Ledig, L.~Theis, F.~Huszar, J.~Caballero, A.~Cunningham, A.~Acosta,
  A.~Aitken, A.~Tejani, J.~Totz, and Z.~Wang, \enquote{Photo-realistic single
  image super-resolution using a generative adversarial network,}
  {\protect\JournalTitle{IEEE Computer Society}}  (2016).

\bibitem{2018Efficient}
Z.~Pengcheng, S.~L. Resendez, R.~R. Jose, J.~C. Jimenez, S.~Q. Neufeld,
  G.~Andrea, F.~Johannes, E.~A. Pnevmatikakis, G.~D. Stuber, and H.~a. Rene,
  \enquote{Efficient and accurate extraction of in vivo calcium signals from
  microendoscopic video data,} {\protect\JournalTitle{Elife}} \textbf{7}
  (2018).

\end{thebibliography}

\end{document}